\documentclass[aps,twocolumn,showpacs]{revtex4}

\usepackage{graphicx}

\begin{document}
\newcommand{\FigureOne}{
	\begin{figure}[t]     
	\includegraphics[width=8.25cm]{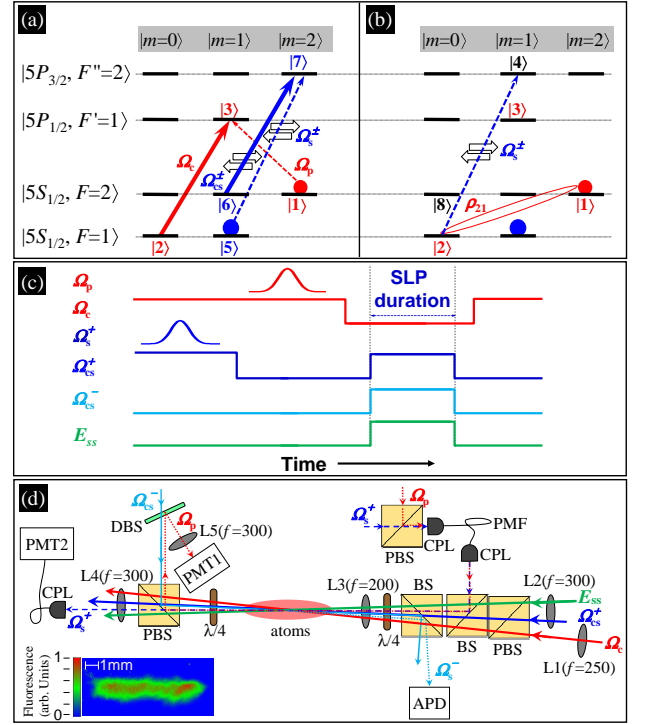}
	\caption{(color online). (a) and (b) Relevant energy levels and laser excitations in the experiment. $\Omega_p$ \& $\Omega_c$ form the 795 nm EIT system of stored light. $\Omega_s^{\pm}$ \& $\Omega_{cs}^{\pm}$ form the 780 nm EIT system of SLP. (c) Timing sequence of stored light switched by stationary light. (d) Schematic experimental setup and image of the atom cloud. PMT1 and PMT2: photo multipliers; APD: avalanche photo detector; PMF: polarization-maintained optical fiber; CPL: optical fiber coupler; BS: cubed beamsplitter; PBS: cubed polarizing beamsplitter; DBS: dichroic beamsplitter; $\lambda/4$: zero-order quarter-wave plate; L1-L5: lenses. }
	\label{fig:TD}
	\end{figure}
}
\newcommand{\FigureTwo}{
	\begin{figure}[t]    
	\includegraphics[width=8.5cm]{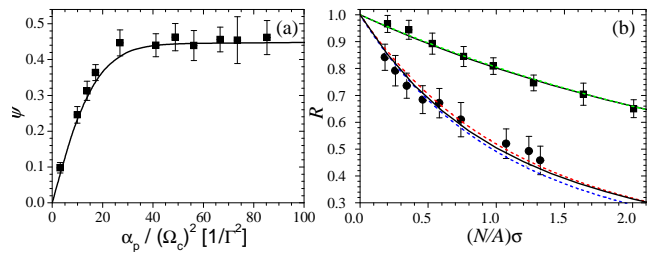}
	\caption{(color online). (a) Switching efficiency versus ratio of OD in the 795 nm EIT system to square of the coupling Rabi frequency in the case of slow light switched by speed-of-c light. $\alpha_p = 3$$\sim$50 and $\Omega_c = 0.77$$\sim$$0.92\Gamma$. (b) Probe attenuation versus number of switching photons per atomic absorption cross section in the two cases of slow light switched by speed-of-$c$ light (squares) and stored light switched by stationary light (circles). In the first case, $\alpha_p = 70$ and $\Omega_c = 0.73\Gamma$; in the second case, $\alpha_p = 72$, $\Omega_c = 0.73\Gamma$, $\alpha_s = 110$, $\Omega^{\pm}_{cs} = 0.75\Gamma$, and the SLP duration is 5.0 $\mu$s. Black solid lines are the theoretical predictions. Green, Red, and blue dashed lines are the plots of Eq.~(\ref{eqn:GaussianR}) with $\psi$ = 0.45, 1.49, and 1.64.}
	\label{fig:psi_eta}
	\label{fig:R_N}
	\end{figure}
}
\newcommand{\FigureThree}{
	\begin{figure}[t]
	\includegraphics[width=8.5cm]{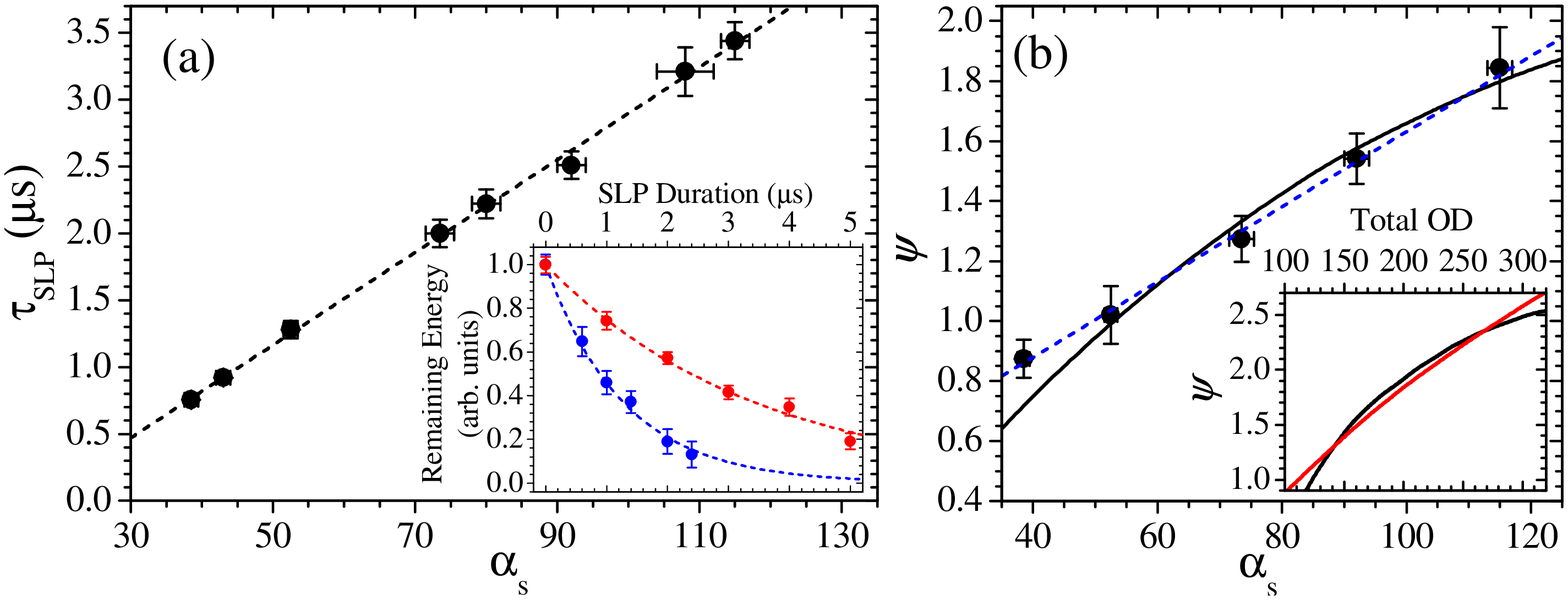}
	\caption{(color online). (a) SLP decay time constant versus OD of the SLP, i.e. the stationary switching pulse. Dashed line is the best fit with a linear function. Inset: SLP remaining energy versus SLP duration at $\alpha_s$ = 115 (red) and 53 (blue). Dashed lines are the best fits with an exponential decay function, showing the time constant = 3.45 and 1.28 $\mu$s. (b) Switching efficiency versus $\alpha_s$. Circles are the experimental data taken under $\alpha_p$ = 68$\sim$72, $\Omega_c$ = 0.70$\sim$0.73$\Gamma$, and $\Omega^{\pm}_{cs}$ = 0.74$\sim$0.76$\Gamma$. Black line is the theoretical prediction calculated at $\alpha_p = 70$, $\Omega_c = 0.72\Gamma$, and $\Omega^{\pm}_{cs} = 0.75\Gamma$. Inset: Total OD is equal to the sum of $\alpha_s$ and $\alpha_p$. Black line is calculated with $\alpha_p = 70$; red line with $\alpha_s = \alpha_p$.}
	\label{fig:tau_OD}
	\label{fig:psi_OD}
	\end{figure}
}
\title{Demonstration of the Interaction between Two Stopped Light Pulses}
\author{Yi-Hsin Chen,$^{1}$ Meng-Jung Lee,$^{1}$ Weilun Hung,$^{1}$
 Ying-Cheng Chen,$^{2,1}$ Yong-Fan Chen,$^{3}$ and Ite A. Yu$^{1}$}
\email{yu@phys.nthu.edu.tw}
\affiliation{
$^{1}$Department of Physics, National Tsing Hua University, Hsinchu 30013, Taiwan \\
$^{2}$Institute of Atomic and Molecular Sciences, Academia Sinica, Taipei 10617, Taiwan \\
$^{3}$Department of Physics, National Cheng Kung University, Tainan 70101, Taiwan
}
\date{October 21, 2011}
\begin{abstract}
We report the first experimental demonstration that two light pulses were made motionless and interacted with each other via a medium. The interaction time is, in principle, as long as possible and a considerable efficiency can be achieved even below single-photon level. We utilized the optical process of one photon pulse switched by another based on the effect of electromagnetically induced transparency to demonstrate the enhancement of optical nonlinear efficiency. With moving light pulses, the switching is activated at energy per area of 2 photons per atomic absorption cross section in the best situation as discussed in [Phys. Rev. Lett. {\bf 82}, 4611 (1999)]. With motionless light pulses, we demonstrated that the switching is activated at 0.56 photons per atomic absorption cross section and that the light level can be further reduced by increasing the optical density of the medium. Our work enters a new regime of low light physics.
\end{abstract}
\maketitle

Efficiency of a nonlinear optical process is equal to the product of transition rate and interaction time. Since the transition rate is dependent on the light intensity, a high intensity laser field is usually required in the nonlinear optics in order to achieve sufficient efficiency. On the other hand, if the interaction time can be made as long as possible, high efficiency can also be achieved even at low-light or single-photon level. With the techniques of light storage (LS) \cite{LsTh,LsNature,LsPRL1,LsPRL2} and stationary light pulse (SLP) \cite{SlpNature,SlpYu09}, we report the first experimental demonstration of enhanced nonlinear efficiency due to long interaction time between two motionless light pulses. In the LS, the storing process converts a light pulse to the ground-state coherence of an ensemble and the retrieving process vice versa. The LS provides a way to transfer quantum states between photons and atoms and can lead to the applications of quantum memory \cite{LsQbit2,LsQbit3,LsQbit4}. The SLP is created by the counter-propagating scheme of four-wave mixing. Unlike the stored light in the LS, an SLP maintains its electromagnetic component which is required for nonlinear optical interactions. Nevertheless, the formation of SLP was observed via the remaining energy released from the medium \cite{SlpNature,SlpYu09}. This work is also the first SLP application and provides the direct evidence of SLPs being electromagnetic fields in the medium. As SLPs significantly increase the interaction time between matter and light, they are very promising for applications in low-light-level nonlinear optics \cite{XpmWithSlp} and studies in quantum many-body physics \cite{SlpBEC,SlpGauge,SlpDirac}.

As a proof-of-principle experiment, the nonlinear optical process that we considered is the all-optical switching (AOS) proposed by Ref.~\cite{PsTheory}. A weak probe and a strong coupling fields form the $\Lambda$-type configuration of electromagnetically induced transparency (EIT) \cite{EitHarris,EitReview} and, hence, the absorption of the probe field is suppressed. The presence of a weak switching field enables the probe absorption owing to the third-order susceptibility \cite{PsYu08,PsPCF}. If one photon switched by another is achievable, the single-photon AOS can lead to the interesting idea of three-fold entangled state and provide applications in quantum information manipulation \cite{PsTheory,PsScience05,PsByPhase}. However, Ref.~\cite{LllTheory} predicts that the switching efficiency has a limit that can hinder the single-photon AOS. The switching efficiency, $\psi$, is defined as
\begin{equation}
	R = \exp[-\psi (N/A) \sigma],
\end{equation}
where $R$ is the probe attenuation, $(N/A)$ is the number of switching photons per unit area, $\sigma = [3\lambda^2/(2\pi)]C^2$ is the atomic absorption cross section, $\lambda$ is the wavelength, and $C$ is the Clebsch-Gordan coefficient of the switching transition. With the probe pulse being slow light and the switching pulse propagating at the speed of $c$, the maximum $\psi$ is 0.5, i.e. the attenuation of e$^{-1}$ requires the light level of 2 photons per $\sigma$ \cite{LllTheory}. The limitation of $\psi$ is resulted from the temporal overlap or interaction time between the two pulses in the medium. A slower probe pulse enhances the third-order susceptibility but shortens the interaction time due to a larger mismatch of the two pulse speeds and does not improve $\psi$ at all.

In order to prolong the interaction time between two light pulses and enhance the switching efficiency, the probe pulse was stopped in the form of ground-state coherence as the stored light and the switching pulse was stopped in the electromagnetic form as the SLP. During the storage of the probe pulse, the stationary switching pulse also drove the transition that destroyed the ground-state coherence of the stored probe pulse. Consequently, the retrieved probe pulse was attenuated due to the presence of the switching pulse. A very long interaction time of 6.9 $\mu$s was achieved by a large optical density (OD) of 190 in our system. This interaction time makes the system analogous to the scheme of trapping light pulses by an optical cavity with a decay time constant of 3.45 $\mu$s or a $Q$ factor of $8\times10^9$. With such interaction time, we achieved about 4-fold improvement in the switching light level as compared with the scheme in Ref.~\cite{LllTheory}. The theoretical study in Ref.~\cite{XpmWithSlp} claims that the nonlinear efficiency due to two motionless light pulses depends on OD linearly. Although our experimental approach is different, we still demonstrated that the interaction time as well as $\psi$ is linearly proportional to OD. In principle, because the interaction time can be as long as possible, $\psi$ has no upper limit with the motionless light pulses. This makes single-photon AOS as well as cross-phase modulation more feasible \cite{XpmWithSlp,XpmYu06}.

To make the two pulses interact with each other, we created two EIT systems in the medium of $^{87}$Rb atoms as shown in Fig.~\ref{fig:TD}(a). Using the method of optical pumping, we put the population only in the two ground states of $|1\rangle$ and $|5\rangle$. With the population in $|1\rangle$, the probe pulse ($\Omega_p$ denotes its Rabi frequency) and coupling field ($\Omega_c$) form the 795 nm EIT system; with that in $|5\rangle$, the switching pulse ($\Omega^{\pm}_s$) and its coupling field ($\Omega^{\pm}_{cs}$) form the 780 nm EIT system. The timing sequence is depicted in Fig~\ref{fig:TD}(c). We first fired $\Omega^+_s$ and stored it by turning off $\Omega^+_{cs}$. Then, we sent $\Omega_p$ and stored it by switching off $\Omega_c$. The two storing processes occurred when the two pulses arrived approximately in the center of the atom cloud. During the storage, we simultaneously turned on counter-propagating $\Omega^+_{cs}$ and $\Omega^-_{cs}$ to produce the stationary switching pulse ($\Omega^{\pm}_s$). At the same time, the double-switching field ($E_{ss}$) was also switched on. Hence, the stored $\Omega_p$ was destroyed by the presence of $\Omega^{\pm}_s$ as illustrated in Fig.~\ref{fig:TD}(b). Finally, we released the remaining $\Omega_p$ from the atom cloud and measured the retrieved probe energy. We applied $E_{ss}$ for the following reason. The switching transition of $|2\rangle \rightarrow |4\rangle$ shown in Fig.~\ref{fig:TD}(b) is also linked by the coupling transition of $|8\rangle \rightarrow |4\rangle$, forming the EIT configuration that prohibits the switching effect. In order to overcome this EIT configuration, $E_{ss}$ drove the transition of $|8\rangle \rightarrow |3\rangle$ and enabled the three-photon transition of $|2\rangle \rightarrow |4\rangle \rightarrow |8\rangle \rightarrow |3\rangle$. It had a Rabi frequency much greater than that of the coupling transition, making the three-photon transition equivalent to the switching transition existing alone \cite{PsTheory}.

\FigureOne

We made the theoretical predictions by numerically solving the Maxwell-Schr\"{o}dinger equations of the light pulses and the optical Bloch equations of the atomic density-matrix operator. The equations for the 780 nm EIT system are given by \cite{SlpYu09,SlpHighOrder}
\begin{eqnarray}   
    \frac{1}{c}\frac{\partial}{\partial t} \Omega^+_s
        + \frac{\partial}{\partial z} \Omega^+_s
        =  i \frac{\alpha_s}{2L} \Gamma \rho^+_{75}, \\
    \frac{1}{c}\frac{\partial}{\partial t} \Omega^-_s
        - \frac{\partial}{\partial z} \Omega^-_s + i \Delta_k \Omega^-_s
        =  i \frac{\alpha_s}{2L} \Gamma \rho^-_{75}. \\
	\frac{\partial}{\partial t} \rho^+_{75}
        = \frac{i}{2} \Omega^+_{s} +\frac{i}{2} \Omega^+_{cs} \rho_{65}
        - \frac{\Gamma}{2} \rho^+_{75}, \\
    \frac{\partial}{\partial t} \rho^-_{75}
       = \frac{i}{2} \Omega^-_{s} +\frac{i}{2} \Omega^-_{cs} \rho_{65}
       - \frac{\Gamma}{2} \rho^-_{75}, \\
    \frac{\partial}{\partial t} \rho_{65}
        = \frac{i}{2} (\Omega^+_{cs})^* \rho^+_{75}
		+ \frac{i}{2} (\Omega^-_{cs})^* \rho^-_{75} - \gamma_s \rho_{65},
\end{eqnarray}
where $+/-$ indicates the fields and optical coherences ($\rho_{ij}$) in the forward/backward direction, $\Gamma$ is the spontaneous decay rate of the excited state, $L$ is the length of the medium, and $\alpha_s$ and $\gamma_s$ denote the OD and the dephasing rate in the 780 nm EIT system. We also included the effect of phase mismatch ($\Delta_k$) \cite{PMM}, occurring in the four-wave mixing process that produces the SLP. Besides the attenuation due to the motion of pulse broadening and the energy loss due to leaking out of the medium cause the SLP decay \cite{SlpDecay}, a larger $\Delta_k L$ makes the decay faster. Furthermore, the equations for the 795 nm EIT system under the presence of the stationary switching field are given by \cite{PsYu08}
\begin{eqnarray}   
	\frac{1}{c}\frac{\partial}{\partial t} \Omega_p
		+ \frac{\partial}{\partial z} \Omega_p
		= i \frac{\alpha_p}{2L} \Gamma \rho_{31}, \\
	\frac{\partial}{\partial t} \rho_{31} = \frac{i}{2} \Omega_p
		+ \frac{i}{2} \Omega_c \rho_{21} - \frac{\Gamma}{2} \rho_{31}, \\
	\frac{\partial}{\partial t} \rho^+_{41}
		= \frac{i}{2} \Omega^+_s \rho_{21} - \frac{\Gamma}{2} \rho^+_{41}, \\
	\frac{\partial}{\partial t} \rho^-_{41}
		= \frac{i}{2} \Omega^-_s \rho_{21} - \frac{\Gamma}{2} \rho^-_{41}, \\
	\frac{\partial}{\partial t} \rho_{21} = \frac{i}{2} (\Omega_c)^* \rho_{31}
		+ \frac{i}{2} (\Omega^+_s)^* \rho^+_{41}
		+ \frac{i}{2} (\Omega^-_s)^* \rho^-_{41} \nonumber \\
		- \gamma_p \rho_{21},
\end{eqnarray}
where $\alpha_p$ and $\gamma_p$ denote the OD and the dephasing rate in the 795 nm EIT system. To study the AOS, we solved Eqs.~(2)-(11) altogether.

The experiment was carried out in the cigar-shaped cloud of cold $^{87}$Rb atoms \cite{CigarMotYu}. Typically, we produced $1.0\times10^9$ atoms with a temperature of about 300 $\mu$K \cite{AtomTempYu}. Figure~\ref{fig:TD}(d) shows the schematic experimental setup and the image of the atom cloud. We performed the measurements of slow light, storage of light, and SLP in the two EIT systems to identify the condition of the experimental system. The slow-light and storage-of-light data demonstrate that the total OD is 190 and the coherence time is around 50 $\mu$s in our system. This large OD was achieved with the technique of dark compressed magneto-optical trap \cite{DarkCMOT1,DarkCMOT2}. The coherence time is such long that the advantage of motionless pulses at the present OD can be fully utilized and the dephasing rates are negligible. The SLP data indicate that $\Delta_k L$ (the product of phase mismatch and medium length) is about 5.5 in our system, which can be caused by misalignment and focusing of the laser beams. This $\Delta_k L$ makes the SLP decay about twice faster and will be used in the calculation thereafter. Details of the experimental method and figures showing the data of the above measurements can be found in Supplemental Material \cite{SSM}.

To ensure that the probe and switching beams completely overlap in the interaction region, we sent these two light beams through the same polarization-maintained optical fiber as shown in Fig.~\ref{fig:TD}(d). Coming out of the optical fiber, the two beams had the same Gaussian intensity profile with a beam waist of $w$. Hence, the probe attenuation should be revised to
\begin{equation}
\label{eqn:GaussianR}
	R = \frac{\int {\rm e}^{-\psi (N/A) \sigma {\rm e}^{-2r^2/w^2}}
		{\rm e}^{-2r^2/w^2} 2\pi r dr}{\int {\rm e}^{-2r^2/w^2} 2\pi r dr}
		= \frac{1-{\rm e}^{-\psi (N/A) \sigma }}{\psi (N/A) \sigma},
\end{equation}
where $(N/A)$ is the number of photons per area corresponding to the center switching intensity. Throughout the experiment, the energy of the input probe pulse was about 8 fJ and that of the input switching pulse ranged between 7 and 150 fJ.

We first verified that the switching efficiency ($\psi$) has a limit in the AOS scheme studied in Ref.~\cite{LllTheory} and used this limit to calibrate the switching intensity. The probe pulse was the slow light. We removed the population from the 780 nm EIT system to make the switching pulse propagate at the speed of $c$. The two pulses were fired at the same time and had the same temporal width. In this AOS scheme, $\psi$ is a function only depending on the ratio of the medium length to the probe spatial width inside the medium \cite{LllTheory}. This ratio is approximately equal to $(\alpha_p\Gamma/\Omega_c^2) /T_p$, where $\alpha_p$ is the OD and $T_p$ is the pulse temporal width. With the input pulse width fixed, Fig.~\ref{fig:psi_eta}(a) shows that $\psi$ monotonically increases with $\alpha_p/\Omega_c^2$ and asymptotically approaches to the maximum value of 0.45. The experimental data (squares) are in good agreement with the theoretical prediction (solid line) calculated from Eqs.~(7)-(11). The behavior of $\psi$ is also very similar to the prediction in Ref.~\cite{LllTheory}. A small difference between $\psi_{\rm max} = 0.45$ obtained here and $\psi_{\rm max} = 0.5$ in the ideal condition should be due to the assumption used in the analytical derivation of Ref.~\cite{LllTheory}. Under the condition that $\alpha_p/\Omega_c^2$ is large enough such that $\psi$ reaches the maximum value, we calibrated the switching intensity ($\propto N/A$) by comparing the experimental data with the theoretical prediction. Because $\psi_{\rm max}$ is affected by the OD and coupling Rabi frequency very little, this made our calibration robust and accurate.

\FigureTwo

We demonstrated that the switching efficiency ($\psi$) in the AOS scheme with the motionless light pulses can far exceed the above limit. The circles in Fig.~\ref{fig:R_N}(b) show such experimental data. The probe attenuation is determined by the ratio of the retrieved probe energies with to without the presence of the switching pulse. The $(N/A) \sigma$ is the switching photon number per atomic absorption cross section at the very beginning of the SLP formation. We determined $(N/A) \sigma$ by considering the temporal profile and calibrated intensity of the input switching pulse as well as the attenuation during the pulse propagation. The theoretical prediction (black solid line) was calculated by numerically solving Eqs.~(2)-(11) and integrating the solutions over the Gaussian beam profile. From the theoretical calculation, we find $\psi$ slightly depends on $(N/A) \sigma$ in the case of stored light switched by SLP. The red and blue dashed lines are the plots of Eq.~(\ref{eqn:GaussianR}) with $\psi = 1.49$ and 1.64, respectively. As a comparison, the squares in Fig.~\ref{fig:R_N}(b) show the experimental data taken in the AOS scheme with the moving pulses as discussed in Ref.~\cite{LllTheory} when $\psi$ reaches the limit. The black solid and green dashed lines are the theoretical prediction and the plot of Eq.~(\ref{eqn:GaussianR}) with $\psi = 0.45$. The comparison between the data of circles and squares clearly manifests the significant enhancement by the AOS scheme with the motionless light pulses. Representative data of signal versus time in the two AOS schemes can be found in Supplemental Material \cite{SSM}.

\FigureThree

Finally, we want to demonstrate that the switching efficiency ($\psi$) in the AOS scheme with motionless light pulses is not limited to the present result and can be further improved by increasing the OD of the system. In Fig.~\ref{fig:tau_OD}(a), the inset shows that the SLP remaining energy decays exponentially against time. The main plot displays the experimental observation that the SLP decay time constant ($\tau_{SLP}$) is linearly proportional to the OD for SLPs ($\alpha_s$). Such result is expected. As mentioned before, the SLP decay is mainly due to broadening motion and leaking out of the medium. A larger OD certainly reduces these decay effects and increases the SLP surviving time. A longer SLP surviving time is equivalent to a longer interaction time between the probe and switching pulses and, hence, results in a better switching efficiency as experimentally demonstrated by the circles in Fig.~\ref{fig:tau_OD}(b). We took the experimental data by fixing the OD for the stored probe pulse ($\alpha_p$) and varying $\alpha_s$. The SLP duration was set to twice of $\tau_{SLP}$. We fitted the experimental data, representatively shown by Fig.~\ref{fig:R_N}(b), with Eq.~(12) to extract $\psi$. The maximum achievable $\psi$ is around 1.8 in this study. The black solid line in Fig.~\ref{fig:tau_OD}(b) is the theoretical prediction calculated from Eqs.~(2)-(11). The agreement between the prediction and data is satisfactory except the data point at the smallest $\alpha_s$. A smaller phase mismatch ($\Delta_k L$) at this $\alpha_s$ causes the discrepancy. The slope of the black line decreases as $\alpha_s$ increases. Such behavior is more noticeable in the inset which covers $\alpha_s$ up to 250. As $\alpha_p$ is kept the same and $\alpha_s$ increases, the spatial profile of the stored probe pulse is unchanged and that of the stationary switching pulse becomes narrower in the medium. The portion of the stored probe pulse that is affected by high switching intensity becomes less. This makes the switching less efficient. By setting $\alpha_p$ equal to $\alpha_s$, the red line in the inset of Fig.~\ref{fig:tau_OD}(b) shows a constant slope. Hence, the switching efficiency can be enhanced by the total OD linearly.

One can increase the atomic density and shorten the medium length ($L$), e.g. cold atoms in an optical dipole trap or Bose condensates, to keep OD high and make the effect of phase mismatch ($\Delta_k L$) negligible. This immediately improves the switching efficiency ($\psi$). For example, if we can make $\Delta_k L \leq 0.5$, $\psi$ will be about 3 at the total OD of 190. Furthermore, the current scheme can be readily applied to the cross-phase modulation (XPM), i.e. the phase of a photon pulse modulated by another. While the probe attenuation is written as Eq.~(1) at $\Delta$ (switching detuning) = 0, the probe phase shift is given by $\psi (N/A) \sigma \Delta \Gamma / (\Gamma^2 +4\Delta^2)$ \cite{XpmYu06}. At $\psi = 3$ and the optimum $|\Delta|$ of $0.5\Gamma$, a single photon tightly focused to an area of around 100$\lambda^2$ can induce a phase shift of the order of 0.01 radians which is certainly measurable \cite{YFC2011PRA}. The single-photon XPM is not far from reality with the current scheme.

In conclusion, we have demonstrated that the switching is activated at 0.56 photons per atomic absorption cross section or the switching efficiency is 1.8 in the scheme of stored light switched by stationary light. We also showed the great potential of this scheme that the switching efficiency is not limited to the present result and can be further improved by increasing the OD of the medium. Our work advances the technology in low-light-level nonlinear optics and quantum information manipulation utilizing photons.

\section*{ACKNOWLEDGEMENTS}
This work was supported by the National Science Council of Taiwan under Grant No. 99-2628-M-007-004. IAY is grateful to S. E. Harris for inspiring discussions. Authors thank S. E. Harris and R. K. Lee for the comments on the manuscript.


\newpage

\newcommand{\cb}{}
\newcommand{\bib}[2]{[#2]}
\setcounter{figure}{0}
\renewcommand{\thefigure}{S\arabic{figure}}
\newcommand{\myref}[2]{\ref{#1}(#2)}
\newcommand{\reffigA}[1]{1(#1)}

\newcommand{\FigureSOne}[1]
{
	\begin{figure}[#1]
	\includegraphics[width=8.5cm]{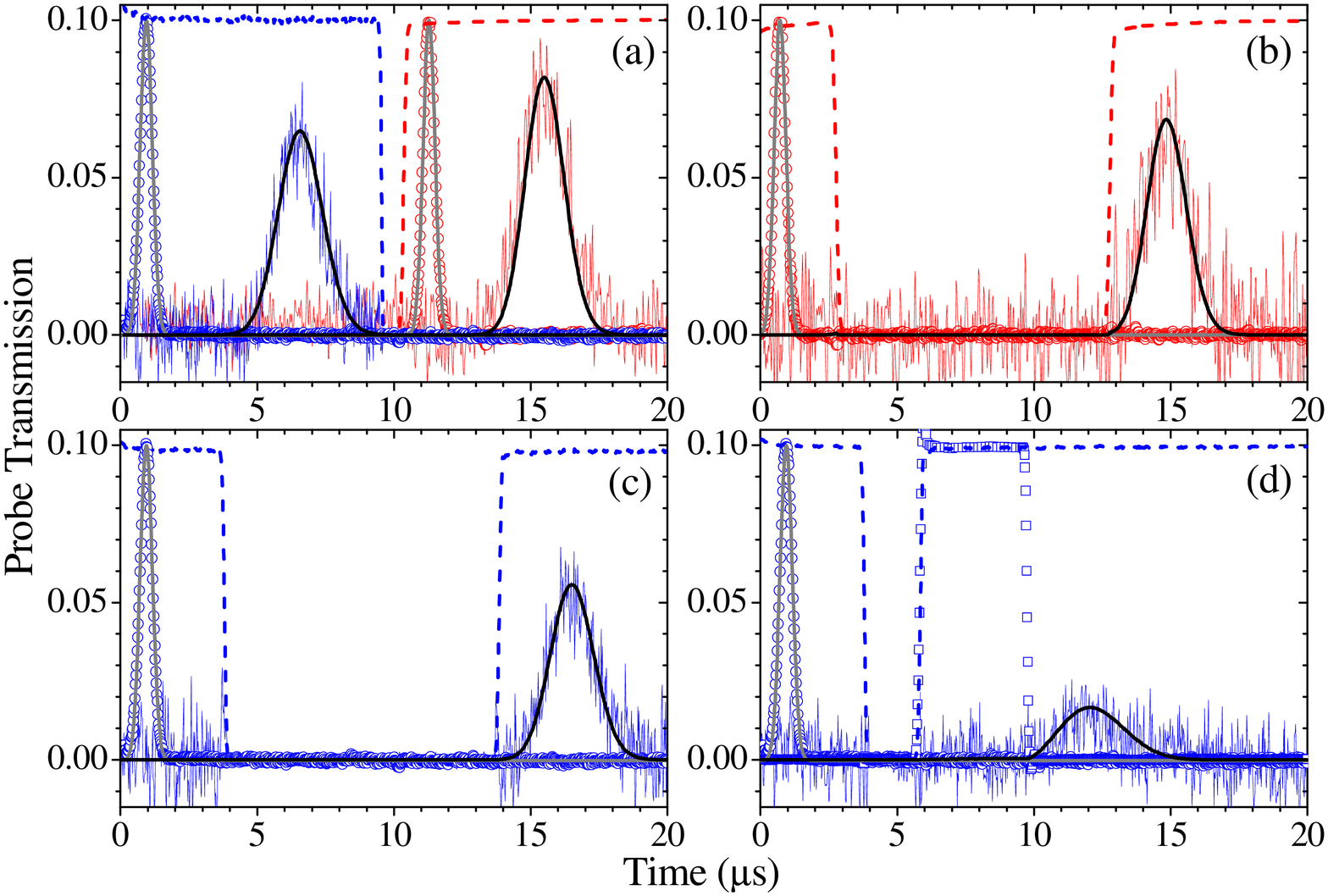}
	\caption{Experimental data and theoretical predictions of the two slow light pulses in (a), storage and retrieval of 795 nm light in (b) as well as of 780 nm light in (c), and stationary light in (d). Red (blue) circles and solid line are the experimental data of the 795 nm (780 nm) input and output probe (switching) pulses. The input pulses are scaled down by a factor of 0.1.
Red dashed line (blue dashed line and open squares) is the experimental data of the 795 nm coupling field (780 nm forward and backward coupling fields). Gray and black solid lines are the Gaussian fits of the input pulses and the theoretical predictions of the output pulses. In the 795 nm EIT system, $\alpha_p = 80$, $\Omega_c = 0.71\Gamma$, and $\gamma_p = 2.5\times10^{-4}\Gamma$. In the 780 nm EIT system, $\alpha_s = 110$, $\Omega^{\pm}_{cs} = 0.73\Gamma$, $\gamma_s = 3\times10^{-4}\Gamma$, and $\Delta_k L = 5.5$.}
	\label{fig:4EITdata}
	\end{figure}
}
\newcommand{\FigureSTwo}[1]
{
	\begin{figure}[#1]
	\includegraphics[width=8.5cm]{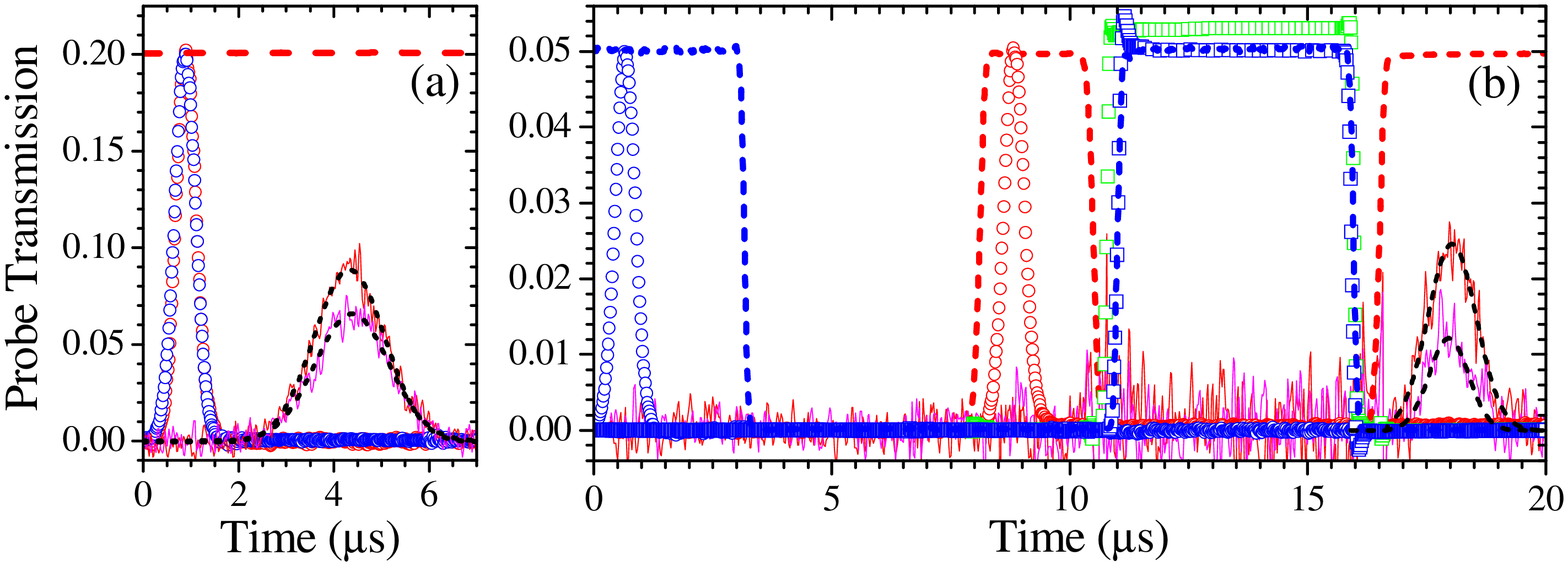}
	\caption{(a) and (b) are the representative data of slow light switched by speed-of-$c$ light and stored light switched by stationary light. The energy per unit area for performing a switch off is about 1 photon per atomic cross section. The data were taken under $\alpha_p = 70$, $\Omega_c = 0.73\Gamma$ in (a) and $\alpha_p = 72$, $\Omega_c = 0.73\Gamma$, $\alpha_s = 110$, and $\Omega^{\pm}_{cs} = 0.75\Gamma$ in (b). In addition to the legends having been described in Fig.~{\cb\ref{fig:4EITdata}}, green open squares are the double-switching field; red and magenta solid lines are the output probe pulses without and with the presence of the switching pulse; black dashed lines are the Gaussian best fits. The input pulses are scaled down by factors of 0.2 and 0.05.}
	\label{fig:2PS}
	\end{figure}
}
\section*{Supplemental Material}

We carried out the experiment in the cigar-shaped cloud of cold $^{87}$Rb atoms produced by a magneto-optical trap (MOT) \bib{CigarMotYu}{26}. The image of the atom cloud is shown in Fig.~\reffigA{d}. Typically, we trapped $1.0\times10^9$ atoms in the MOT with a temperature of about 300 $\mu$K \bib{AtomTempYu}{27}. With the method of optical pumping, we put the population only in the two ground states of $|F=2,m=2\rangle$ ($|1\rangle$) and $|F=1,m=1\rangle$ ($|5\rangle$) as shown in Fig.~\reffigA{a}. Let $F$, $F'$, and $F''$ represent the hyperfine levels in the $5S_{1/2}$, $5P_{1/2}$, and $5P_{3/2}$ states, where the spontaneous decay rate ($\Gamma$) of the $F'$ and $F''$ levels is approximately $2\pi\times 6$ MHz. In the 795 nm EIT system, the probe (whose Rabi frequency is denoted as $\Omega_p$) and coupling ($\Omega_c$) fields were circularly polarized with left ($\sigma_-$) and right ($\sigma_+$) helicities. The two fields drove the transitions of $|F=2,m=2\rangle$ $\rightarrow$ $|F'=1,m=1\rangle$ and $|F=1,m=0\rangle$ $\rightarrow$ $|F'=1,m=1\rangle$. They came from two lasers, one of which was injection-locked by another. In the 780 nm EIT system, the switching ($\Omega^+_s$, i.e. the probe in the EIT point of view) and forward/backward coupling ($\Omega^{\pm}_{cs}$) fields all had the $\sigma_+$ polarization and drove the transitions of $|F=1,m=1\rangle$ $\rightarrow$ $|F''=2,m=2\rangle$ and $|F=2,m=1\rangle$ $\rightarrow$ $|F''=2,m=2\rangle$. The switching and two coupling beams came from the diode lasers, which were all phase-locked to each other. The double-switching ($E_{ss}$) field drove the transition of $|F=2, m=0\rangle$ $\rightarrow$ $|F'=1, m=1\rangle$ and its purpose has been described in the paper. The experimental setup is illustrated in Fig.~\reffigA{d}. To ensure that the probe and switching beams completely overlap in the interaction region, we sent these two light beams through the same polarization-maintained optical fiber. Coming out of the optical fiber, they were focused into a beam size with the e$^{-2}$ full width of 150 $\mu$m and propagated along the major axis of the atom cloud. A small DC magnetic field of about 0.1~G was applied. All laser beams used in the experiment were independently switched by acoustic-optic modulators. 

Before performing each measurement, we momentarily employed the technique of dark compressed MOT (dark-CMOT) for about 7 ms to increase the atomic density of the system \bib{DarkCMOT1,DarkCMOT2}{28, 29}. The transverse (longitudinal) magnetic field gradient of the MOT was increased from 11 (1.3) to 14 (1.7) G/cm and the intensity of the MOT repumping field was reduced from 1.5 mW/cm$^2$ to 4.4 $\mu$W/cm$^2$. This dark-CMOT process enhanced the optical density (OD) of the system by more than 3 folds. The maximum OD of about 190 was achieved in this work. Denote the time of firing the switching pulse as $t = 0$. After the dark-CMOT process, we turned off the magnetic field at $t = -700$ $\mu$s. The 795 nm coupling and 780 nm forward coupling fields were switched on at $t = -50~\mu$s. Only the ground states of $|1\rangle$ and $|5\rangle$ were not driven by these two coupling fields. This is how we can accumulate all population in these two states. The population can be pumped out of $|1\rangle$ by the MOT trapping field. We adjusted the switching-off time of the trapping field to control the populations in the two states and, consequently, the ODs in the 795 nm ($\alpha_p$) and 780 nm ($\alpha_s$) EIT systems. For example, switching off the trapping field at $t = -20$ $\mu$s results in the ODs in Fig.~\myref{fig:2PS}{b}. Once the populations or ODs in $|1\rangle$ and $|5\rangle$ were prepared, we proceeded a designated measurement.

\FigureSOne{b}

To identify the experimental condition, we performed the measurements of slow light, storage of light, and stationary light. Figure~\myref{fig:4EITdata}{a} shows the experimental slow-light data of the 795 nm probe and 780 nm switching pulses taken in the same measurement. From the delay times and attenuations of the output pulses, we can estimate the ODs and the coupling Rabi frequencies in the two systems. The measurement shows that the total OD ($= \alpha_p + \alpha_s$) in the experiment is 190. In Fig.~\myref{fig:4EITdata}{b}, we stored the probe pulse in the atoms by switching off the coupling field; in Fig.~\myref{fig:4EITdata}{c}, we stored the switching pulse in the similar way. The stored pulse was retrieved by switching on the coupling field. After a 10-$\mu$s storage time, the remaining energies of the pulses released from the atoms determine the dephasing rate of the probe pulse ($\gamma_p$) and that of the switching pulse ($\gamma_s$). The dephasing rates show that the coherence time in our system is around 50 $\mu$s. In Fig.~\myref{fig:4EITdata}{d}, the timing sequence is the following. The switching pulse was first stored. During the storage, we simultaneously switched on $\Omega^+_{cs}$ and $\Omega^-_{cs}$ to form the stationary light pulse (SLP). After a 4-$\mu$s duration, the SLP was converted to the moving pulse leaving the atoms by shutting off only $\Omega^-_{cs}$. Details of the SLP formation can be found in Ref.~\bib{SlpYu09}{6}. All the solid black lines in the figures are the theoretical predictions calculated from Eqs.~(2)-(11). The comparison between the SLP data and the prediction indicates that there existed $\Delta_k L = 5.5$ in our system, where $L$ is the medium length and $\Delta_k$ is the phase mismatch \bib{PMM}{24}, occurring in the four-wave mixing (FWM) process that produces the SLP. The experimental data and the theoretical predictions are in good agreement, providing the accurate experimental condition.

\FigureSTwo{t}

We used two photo multipliers (Hamamatsu PMT H6780-20 and amplifier C9663 or C6438-01) to detect the signals of the probe and switching pulses. The outputs from the PMTs were averaged 256 times by a digital oscilloscope (Agilent MSO6014A). The peak powers of the output probe and switching pulses in Fig.~\myref{fig:4EITdata}{a} were about 1 nW and 8 nW.

In the study of all-optical switching (AOS), we optimized the frequency of the switching pulse to maximize the switching efficiency. Figure~\myref{fig:2PS}{a} shows the representative data of the slow-light probe pulse switched by the switching pulse propagating at the speed of $c$. We depleted the population in the state of $|5\rangle$ and fired the both pulses at the same time. The both input pulses had the same Gaussian temporal profile with e$^{-2}$ full width of 0.88 $\mu$s. Figure~\myref{fig:2PS}{b} shows the representative data of the stored probe pulse switched by the stationary switching pulse. Refer the timing sequence in Fig.~\reffigA{c}. We first fired the switching pulse and stored it by turning off the 780 nm forward coupling field. Then, we sent the probe pulse and stored it by switching off the 795 nm coupling field. The two storing processes occurred when the two pulses arrived approximately in the center of the atom cloud. During the storage, we simultaneously turned on $\Omega^+_{cs}$ and $\Omega^-_{cs}$  to produce the SLP. At the same time, the double-switching field was also switched on. At the SLP period, e.g. 5.0 $\mu$s used in Fig.~\myref{fig:2PS}{b}, the stored probe pulse was destroyed by the presence of the SLP as illustrated in Fig.~\reffigA{b}. Finally, we released the remaining probe pulse from the atom cloud and measured the probe energy. The far-detuned transition of $|F=2,m=2\rangle \rightarrow |F''=3,m=3\rangle$ induced by $\Omega^{\pm}_{cs}$ reduced the stored probe signal. Hence, the probe attenuation was determined by the ratio of the retrieved probe energies with to without the existence of the switching pulse, while all the other laser fields were always present. The red and magenta solid lines in Figs.~\myref{fig:2PS}{a} and \myref{fig:2PS}{b} are the output probe pulses without and with the presence of the switching pulse.

\end{document}